\documentclass[12pt,preprint]{aastex}
\usepackage{emulateapj5}

\slugcomment{Received 2007 August 22; accepted 2008 April 15}

\shortauthors{Smith et al.}
\shorttitle{PDS: ERGs and Cluster Candidates}

\begin{document}

\title{The Phoenix Deep Survey: Extremely Red Galaxies and Cluster Candidates}

\author{Anthony G. Smith\altaffilmark{1,10},
Andrew M. Hopkins\altaffilmark{1},
Richard W. Hunstead\altaffilmark{1},
Samuel J. Schmidt\altaffilmark{2,3},
Jos{\'e} Afonso\altaffilmark{4,5},
Antonis E. Georgakakis\altaffilmark{6},
Lawrence E. Cram\altaffilmark{7},
Bahram Mobasher\altaffilmark{8} and
Mark Sullivan\altaffilmark{9}
}

\affil{
\begin{enumerate}
 \item School of Physics, University of Sydney, NSW 2006, Australia.
 \item Department of Physics and Astronomy, University of 
Pittsburgh, 3941 O'Hara Street, Pittsburgh, PA 15260.
\item Physics Department, University of California, 1 Shields Avenue, Davis, CA 95616.
\item Observat\'orio Astron\'omico de Lisboa, Faculdade de Ci\^encias, Universidade de Lisboa, Tapada da Ajuda, 1349-018 Lisbon, Portugal.
\item Centro de Astronomia e Astrof\'{\i}sica da Universidade de Lisboa, Lisbon, Portugal.
 \item Imperial College of Science Technology and Medicine, Blackett 
Laboratory, Prince Consort Rd, London SW7 2BZ.
 \item Chancelry 10, The Australian National University, Canberra 
ACT 0200, Australia.
 \item Space Telescope Science Institute, 3700 San Martin Drive, 
Baltimore, MD 21218.
 \item Department of Physics (Astrophysics), University of Oxford, Denys Wilkinson Building, Keble Road, Oxford, OX1 3RH, UK.
 \item Email: asmith@physics.usyd.edu.au
\end{enumerate}
}

\begin{abstract}

We present the results of a study of a sample of 375 Extremely Red 
Galaxies (ERGs) in the Phoenix Deep Survey, 273 of which constitute 
a subsample which is 80\% complete to $K_{s} = 18.5$ over an area of 1160 arcmin$^2$. The angular 
correlation function for ERGs is estimated, and the association of ERGs 
with faint radio sources explored. We find tentative evidence that ERGs 
and faint radio sources are associated at $z \gtrsim 0.5$. A new
overdensity-mapping algorithm has been used to characterize 
the ERG distribution, and identify a number of cluster candidates, including a likely cluster
containing ERGs at $0.5 < z < 1$. Our algorithm is also used in an attempt to probe
the environments in which faint radio sources and ERGs are 
associated. We find limited evidence that the $I - K_{s} > 4$ 
criterion is more efficient than $R - K_{s} > 5$ at selecting dusty 
star-forming galaxies, rather than passively evolving ERGs.

\end{abstract}

\keywords{galaxies: clusters: general -- galaxies: high-redshift -- 
galaxies: structure -- infrared: galaxies -- surveys}

\section{Introduction}

Extremely red objects (EROs) comprise stars, stellar remnants and 
galaxies. EROs that are not galaxies include brown dwarfs and 
protostars, and have been studied (e.g. \citealp{Fan2000}) at 
infrared $K$ magnitudes $\lesssim$ 15. At fainter magnitudes, most EROs 
are galaxies. Use of the term ``ERO" to describe both classes of 
objects is widespread, but to avoid confusion, we refer hereafter to 
the ERO galaxies as extremely red galaxies (ERGs). They have very 
red colours ($R - K_{s} > 5$, $I - K_{s} > 4$), and are thought to 
include members of two distinct populations (e.g.\ 
\citealp{Cim2003}, \citealp{McC2004}, \citealp{Dad2004}): dusty 
star-forming galaxies (including some active galactic nuclei, AGNs) 
and evolved ellipticals at $z \gtrsim 1$. Following the example of 
\citeauthor{Roc2002} \citeyearpar{Roc2002}, we will refer to the 
dusty star-forming ERGs as ``dsfERGs", and to the passively evolving 
ERGs as ``pERGs". The relative mix of the two populations, which is 
a function of magnitude, colour and redshift, has not been well 
constrained in the literature, although progress has been made (e.g.\ 
\citealt{Cim2003}, \citealt{Mou2004}).
 The red colours in dsfERGs are caused by dust absorbing shorter 
wavelength light and in pERGs by old stellar populations.

Different models of galaxy evolution --- monolithic collapse with 
passive luminosity evolution (PLE) versus hierarchical merging in a 
cold dark matter (CDM) cosmology --- predict different properties 
for galaxy populations. These involve differences in the formation 
scenario for ellipticals, and in the evolution of the large-scale 
structure of the universe. The hierarchical merging model predicts 
that the number of ellipticals should increase with time, as they 
form through the merging of spiral or irregular systems.

\citet{Eis2000} found that a high fraction of red galaxies, 
corresponding to redshifts $z > 1$, was at least as consistent with 
PLE as with CDM merger models. Scodeggio \& Silva 
\citeyearpar{Sco2000} found that the space density of ERGs was 
consistent with no change in the volume density of ellipticals from 
$z = 0$ to 1.5. However, Rodighiero et al. \citeyearpar{Rod2001} 
found evidence for CDM merger models in the form of a decrease in 
the density of E/S0 galaxies between $z = 1$ and 1.5, with a 
corresponding drop in the density of ERGs. A characteristic 
limitation of these early ERG surveys was large field-to-field 
variations in ERG density (e.g.\ \citealt{Bar1999} compared with 
\citealt{McC2000}). This is largely a result of cosmic variance, but 
is also related to survey sensitivity and how the ERG class is 
defined (\citealt{McC2004}). ERG density is a strong function of 
both the survey depth and the colour threshold used to define the 
ERG population.

ERGs include galaxies that formed at high redshift and evolved 
passively since then (e.g. \citealt{Fir2002}). For example, Spinrad 
et al.\ \citeyearpar{Spi1997} studied a pERG at $z = 1.55$, 
estimating that it had a formation redshift $z_{f} \geq 5$ and 
identifying it as possibly the oldest galaxy known at $z \gtrsim 1$. 
The detection of highly evolved ellipticals at $1 < z < 2$ by 
Ben{\'{\i}}tez et al. \citeyearpar{Ben1999}, with a corresponding lack of 
luminous blue ellipticals, also implied a formation redshift for 
ellipticals of $z_{f} \geq 5$. From ERGs in the K20 survey, Cimatti 
et al. \citeyearpar{Cim2003} deduce a minimum formation redshift of 
about 2 for pERGs at $z \approx 1$, and a significant range for the 
formation redshift of about 2.2 to 4. 

Even though the evolution of the star-formation rate (SFR) in 
different environments has not been comprehensively constrained 
(e.g.\ \citealt{Pog2006}), it is clear that the lowest SFRs in the 
nearby universe, found in high mass elliptical galaxies, were the 
sites of the highest SFRs in the distant past. Smail et al.\ 
\citeyearpar{Sma1999} suggest that the most extreme ERGs are mainly 
of the dusty star-forming variety, and that studying the dsfERG 
population may tell us much about the nature of star formation in 
the early universe.

Measurement of the densities of pERGs may therefore be used to 
test models of galaxy evolution in terms of the mass assembly of 
galaxies, with the clustering properties of both pERGs and dsfERGs 
being used to test the clustering predictions of these scenarios. 
The importance of ERGs for such tests is emphasized by Georgakakis 
et al. \citeyearpar{Geo2006}, who find that ERGs contribute almost 
half the stellar mass density of the universe at $z \approx 1$. 
Furthermore, both ERG classes may provide valuable information about 
the formation of the earliest galaxies and clusters.
Faint radio sources are also considered to identify 
clusters (e.g. Best \citeyear{Bes2000}, Buttery et al. 
\citeyear{But2003}), but it has not been 
conclusively shown that ERGs and faint radio sources trace similar 
structures at high redshift (Georgakakis et al. \citeyear{Geo2005}). 

In this paper, we present the results of a moderately deep 
near-infrared $K_{s}$ survey (Section 2) covering $\approx 1160$ 
arcmin$^{2}$ in the Phoenix Deep Survey region.  Our survey covers a 
larger area than the $K_{s}$-band survey detailed in Sullivan et 
al.\ \citeyearpar{Sul2004} but to a shallower depth. Our 
corresponding ERG sample, defined in Section 3, is about the same 
numerical size as that studied by Georgakakis et al. 
(\citeyear{Geo2005}, \citeyear{Geo2006}). Our ERG sample is 
complemented by deep radio data, which we use to examine the 
association of ERGs and radio sources. Deep multiwavelength coverage 
allows us to investigate differences in clustering properties which 
may be introduced by different ERG selection criteria. We examine 
the environments and clustering of our sample in Sections 4 and 5, 
focussing on a possible cluster at $0.5 < z < 1$ in Section 6. Where 
necessary in this work, we have adopted a concordance cosmological model with $H_{0} = 70$ km s$^{-1}$ 
Mpc$^{-1}$, $\Omega_{M} = 0.3$ and $\Omega_{\Lambda} = 0.7$. All 
magnitudes are in the Vega system.

\section{The Phoenix Deep Survey} 
The Phoenix Deep Survey\footnote{See also http://www.physics.usyd.edu.au/$\sim$ahopkins/phoenix/} (PDS) is a 
multiwavelength survey aimed at studying the nature and evolution of 
faint radio sources. The radio data, which reaches a minimum $1\,\sigma$ 
noise level of 12 $\mu$Jy at its most sensitive, defines the survey area of
4.56 deg$^2$ in the southern constellation 
Phoenix. The radio observations (Hopkins et al. 
\citeyear{Hop1998}, \citeyear{Hop1999}, \citeyear{Hop2003}) were 
carried out between 1994 and 2001 at the Australia Telescope Compact 
Array (ATCA).

\subsection{Survey Multiwavelength Data Set}
Deep multicolour images are available for part of the PDS in $U$, 
$B$, $V$, $R$ and $I$ bands. \textit{BVRI} data were obtained in 2001 using 
the Wide Field Imager (WFI) at the Anglo-Australian Telescope (AAT), 
Siding Spring Observatory. $U$ band data were collected at the Cerro 
Tololo Inter-American Observatory (CTIO) in 2003, while deep 
near-infrared data ($K_{s} \lesssim 20$) are acquired for a 180 
arcmin$^{2}$ region using the Son of ISAAC (SofI) instrument at the 
European Southern Observatory's (ESO) New Technology Telescope (NTT) 
at La Silla. Details of the optical and near-infrared observations 
are given in Sullivan et al. \citeyearpar{Sul2004}. The $R-$ and 
$I-$ band catalogs have 5\,$\sigma$ limiting magnitudes of 24.61 
and 24.67 respectively, using SExtractor's (\citealt{Ber1996}) MAG\_AUTO photometry.

\subsection{Extended Near-Infrared Dataset}
In addition to the SofI data, near-infrared images ($K_{s} \lesssim$ 18.5) covering an area of 1160 
arcmin$^{2}$ of the PDS were obtained with the Infrared Side Port 
Imager (ISPI) camera at CTIO on 2004 September 25. The conditions 
were photometric, with seeing ranging from $0\farcs8$ to $1\farcs2$. 
We observed 11 ISPI pointings using a random dither pattern, nine 
with a total exposure time of 30\,minutes, and the remaining two 
with exposure times of 21 and 15\,minutes. Image processing was 
performed using standard NOAO IRAF\footnote{IRAF is distributed by 
the National Optical Astronomy Observatory, which is operated by the 
Association of Universities for Research in Astronomy, Inc., under 
cooperative agreement with the National Science Foundation.} 
routines. Dark frames were subtracted from each image, as were 
running sky flats, constructed from medians of up to 10 object 
frames. The IRAF tasks \verb+mscgetcat+ and \verb+msccmatch+ were 
used to accurately register the images for co-addition. The final 
astrometry was done by reference to 2MASS\footnote{This publication 
makes use of data products from the Two Micron All Sky Survey, 
2MASS, which is a joint project of the University of Massachusetts 
and the Infrared Processing and Analysis Center/California Institute 
of Technology, funded by the National Aeronautics and Space 
Administration and the National Science Foundation.} 
(\citealt{Skr2006}) source positions. Residual offsets with respect 
to 2MASS positions, and to fainter PDS sources in $R$-band images, 
were fitted by a Gaussian with $\sigma=0\farcs3$.

\begin{figure}[h]
\centering
\includegraphics[scale=0.5]{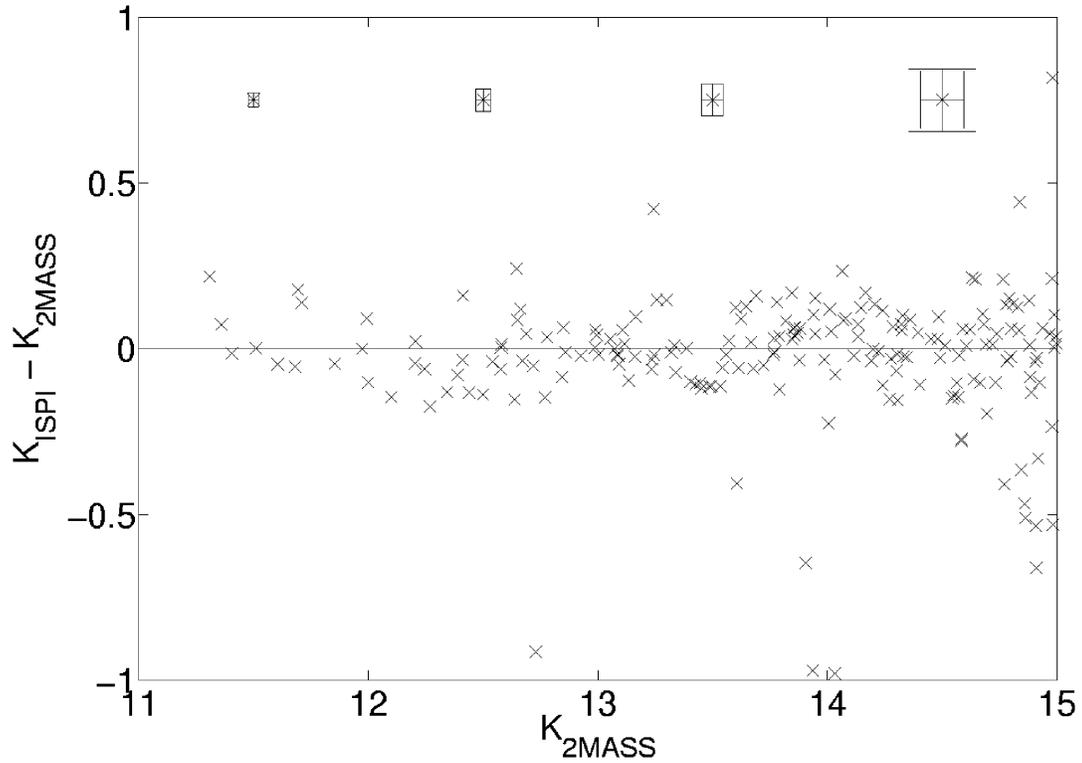}
 \caption{A comparison of the 2MASS and ISPI magnitudes after 
photometric calibration, for objects with $K_{s} < 15$. The points 
significantly below the line arise from extended or saturated
objects, resulting in erroneous integrated fluxes. The error boxes
indicate the typical uncertainties as a function of magnitude.}
 \label{phoenixand2mass}
\end{figure}

We employed SExtractor on each of the eleven 
$K_{s}$-band images using their respective exposure maps as weight 
images, with a detection threshold of 1.5 times the noise level. The low 
signal-to-noise regions on the edges of the co-added images, a 
result of the dithering of individual exposures, are excluded with 
simple right ascension and declination cuts.  The overlap between 
the images was such that we were able to maintain contiguity 
throughout the area of our $K_{s}$-band survey. The final 
catalogue contains 4834 objects.
Within the area 
of our catalogue, there are 230 sources, mostly stars, which also 
have 2MASS identifications. For each individual image, we tied our 
photometry to 2MASS by adjusting the magnitude zeropoints in our 
SExtractor input to give the tightest correlation between ISPI 
and 2MASS magnitudes (see Figure \ref{phoenixand2mass}).

\begin{figure}[h]
\centering
\includegraphics[scale=0.5]{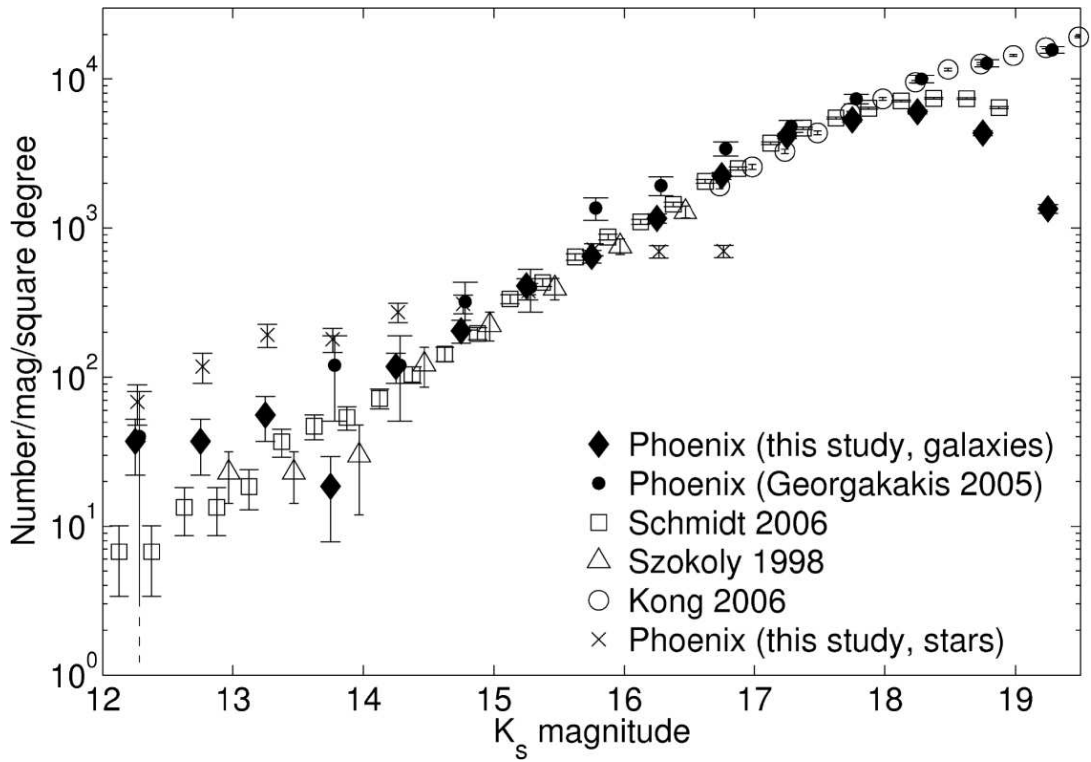}
 \caption{Differential galaxy source counts, compared with counts 
from Georgakakis et al.\ \citeyearpar{Geo2005}, Schmidt et al.\ \citeyearpar{Sch2006}, Szokoly et al.\ 
\citeyearpar{Szo1998} and Kong et al.\ \citeyearpar{Kon2006}.
Differential star counts are also shown. The large error bar for the 
Georgakakis et al.\ point at $K_s=12.25$ arises from a single object in 
that bin. The high counts at the bright end of our sample may arise from a small
systematic overestimation of brightness by SExtractor at 
these levels.}
 \label{galaxycounts}
\end{figure}

Following Schmidt et al. \citeyearpar{Sch2006}, we performed 
star-galaxy separation by comparing SExtractor's fixed aperture 
($3\farcs5$ diameter) magnitude with its MAG\_AUTO magnitude, after 
which we used MAG\_AUTO for all subsequent analysis. Our 
$K_{s}$-band galaxy counts shown in Figure \ref{galaxycounts} are 
consistent with other recent measurements. The completeness of the 
catalogue was estimated to be 80\% to $K_s=18.5$ from comparison 
with source counts from Kong et al.\ (\citeyear{Kon2006}).

\section{The ERG Sample}
\subsection{ERG Selection}
\begin{figure}[h]
\centering
\includegraphics[scale=0.5]{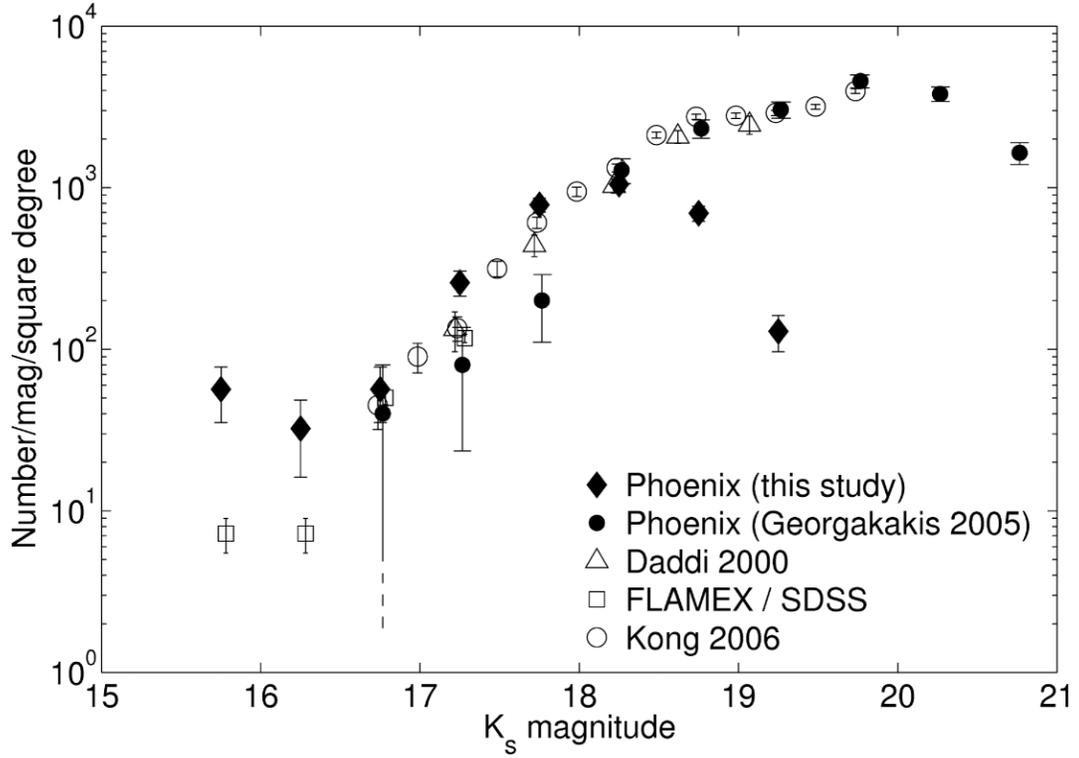}
 \caption{Differential ERG number counts for the $R-K_{s} > 5$-selected sample compared with Georgakakis et al. 
\citeyearpar{Geo2005}, Daddi et al. \citeyearpar{Dad2000}, FLAMEX/SDSS (\citealt{Els2006}) and Kong et al.\ \citeyearpar{Kon2006}.
Our counts at the bright end are subject to small number statistics.}
 \label{ERGnumcounts}
\end{figure}

We matched our near-infrared catalogue with the existing optical 
catalogue (\citealt{Sul2004}), counting objects within 2$''$ as 
matches. We constructed two samples of ERGs based on the two-colour 
criteria: $R - K_{s} > 5$ and $I - K_{s} > 4$. This resulted in a 
sample of 375 $R - K_{s}$-selected ERGs, 273 of which are brighter 
than $K_{s} = 18.5$. Similarly, we derived a sample of 346 $I - 
K_{s}$-selected ERGs, 256 of which are brighter than $K_{s} = $18.5. 
301 ERGs are common to both samples, 228 of which are brighter than 
$K_{s} = $18.5. Figure \ref{ERGnumcounts} shows a comparison of the 
differential $R - K_{s}$ ERG counts with those from previous studies. ERGs with 
more extreme colours close to the $K_{s}$-band detection limit may 
not be detected in $R$-band, and will therefore contribute to the incompleteness 
seen in Figure \ref{ERGnumcounts}.

\subsection{ERG Correlation Function}
To characterize the clustering properties of this ERG sample, we use the two-point angular correlation function.
We follow the prescription detailed by Georgakakis et al.\ \citeyearpar{Geo2005}, using the Landy \& Szalay\ \citeyearpar{Lan1993} estimator for calculating $w(\theta)$, with uncertainties that are assumed to be Poissonian.
We also apply an integral constraint ($C$), again as detailed by Georgakakis et al.\ \citeyearpar{Geo2005}, assuming the correlation function slope $\delta = 0.8$, which gives $C = 5.42$ for our survey area.

\begin{figure}[h]
\centering
\includegraphics[scale=0.25]{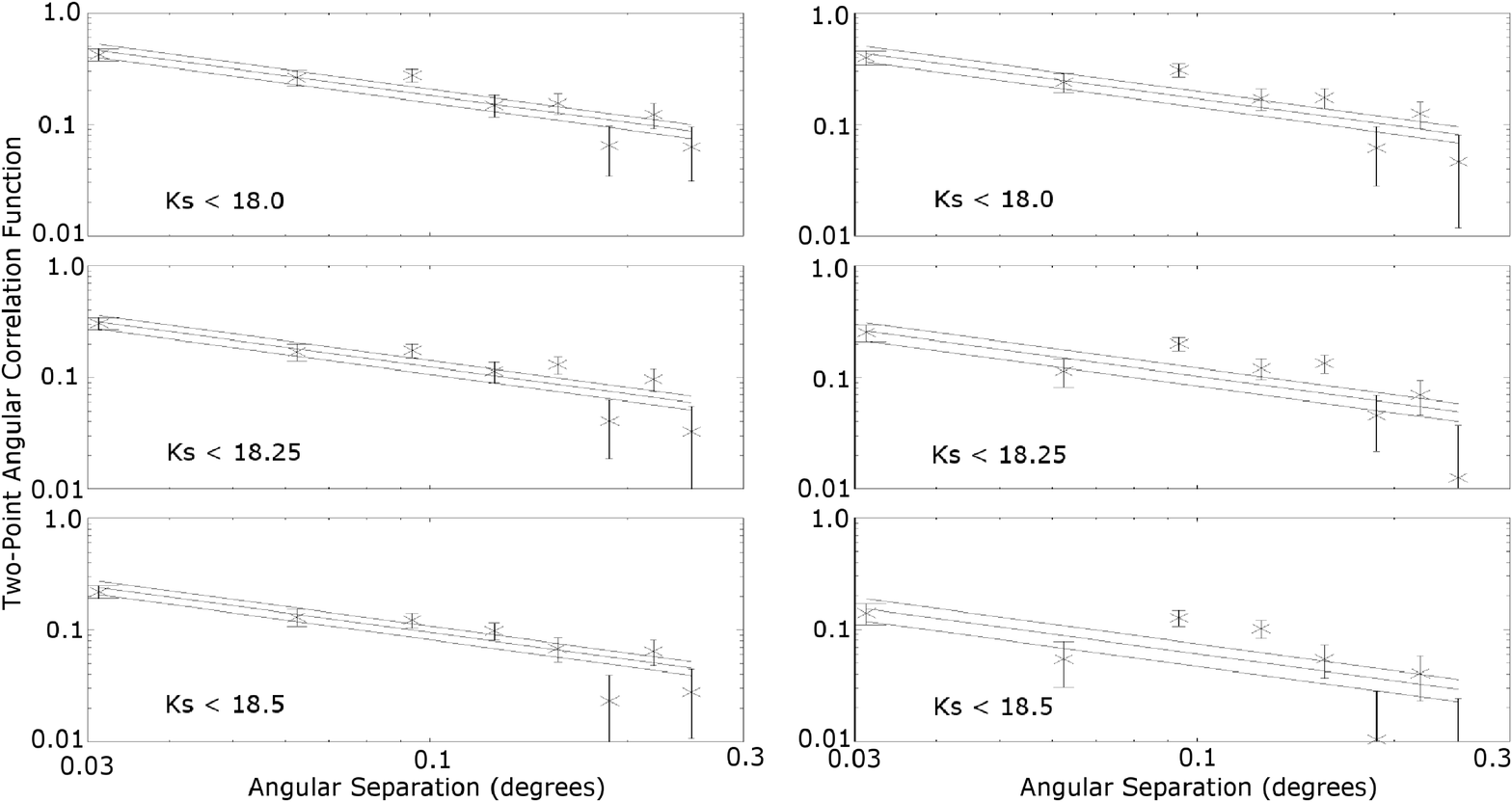}
 \caption{Correlation functions for the ERG samples as a function of 
limiting magnitude.  {\it Left:} $R - K_{s}>5$ and {\it Right:} $I - 
K_{s}>4$.  The value of $\delta$ is fixed at 0.8 and the upper and 
lower lines signify $\pm 1\,\sigma$ uncertainties. The increased 
scatter in the $I - K_{s}$-selected samples, introduced by a small 
number of objects that are not common to both samples, illustrates 
the difference between the clustering properties of the ERG samples 
selected by the two criteria.}
 \label{corfs}
\end{figure}

\begin{figure}[h]
\centering
\includegraphics[scale=0.5]{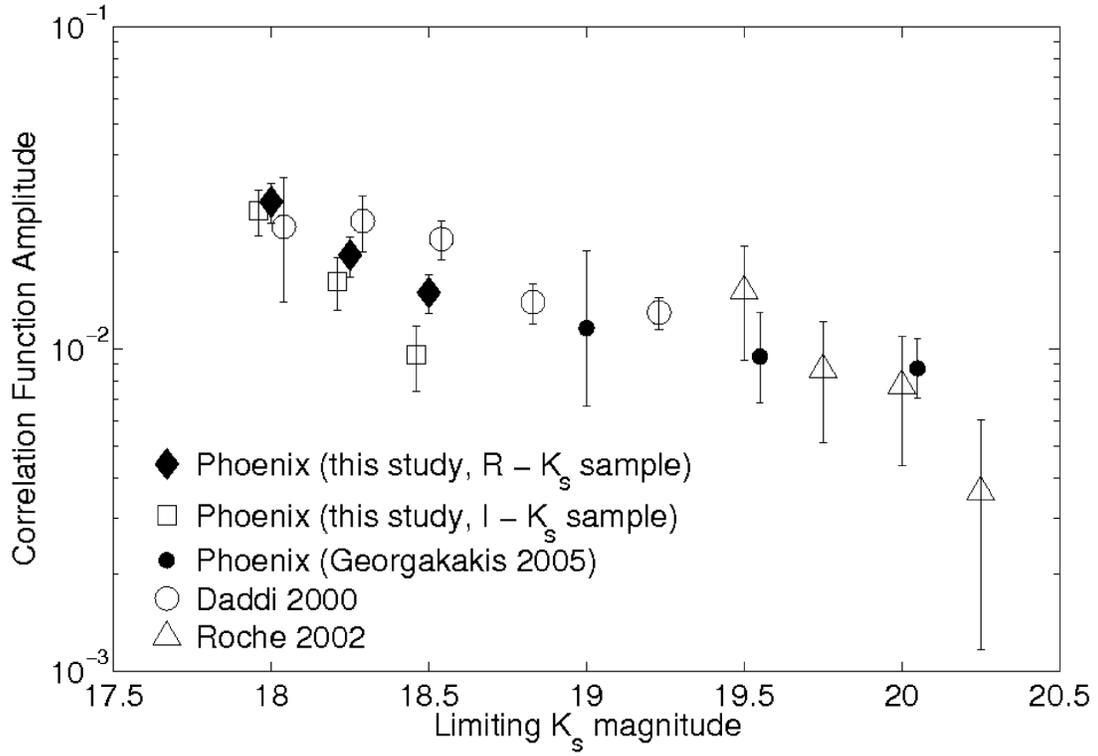}
 \caption{Clustering amplitudes for the $R - K_s>5$ and $I - K_s>4$ 
samples compared with those found by Georgakakis et al.\ 
\citeyearpar{Geo2005}, Daddi et al.\ \citeyearpar{Dad2000} and Roche 
et al.\ \citeyearpar{Roc2002}. For clarity, points at the same 
limiting magnitude have been offset horizontally by $\pm 0.04$ mag. 
At a limiting magnitude of $K_{s} = 18.5$ our clustering amplitudes 
are significantly lower than Daddi et al.\ because of our survey 
incompleteness.}
 \label{clusteringamplitudes}
\end{figure}

In Figure \ref{corfs} we present the correlation functions for the 
ERGs from our data. We find significant clustering for limiting 
$K_{s}$ magnitudes between 18 and 18.5 (Table 
\ref{clusteringamplitudetable}). The clustering amplitudes for our 
ERG sample are consistent with other recent measurements (Figure 
\ref{clusteringamplitudes}), and are much greater than the 
amplitudes which have been measured for all $K_{s}$-band galaxies to the 
same limiting magnitudes (e.g. \citealp{Dad2000}). There appears to 
be a difference between the clustering properties of the ERGs 
selected by our two different criteria. In Table 
\ref{clusteringamplitudetable}, at each limiting magnitude, the 
clustering amplitude and confidence level of the clustering signal 
for ERGs selected using the $R - K_{s}$ criterion is greater than 
that found using $I - K_{s}$ selection. This is also seen in Figure 
\ref{corfs}, supporting the suggestion (V{\"a}is{\"a}nen \& 
Johansson \citeyear{Vai2004}) that $I - K_{s}$ selection may be more 
sensitive to the dsfERG population, as these galaxies are expected 
to cluster less strongly than the pERGs.
We stress that differences in the redshift distributions of ERGs selected by the different colour criteria are likely to affect their measured clustering properties.
Observed-frame colour tracks given by McCarthy \citeyearpar{McC2004} show that the $R - K_{s}$ criterion should, in principle, be more efficient at selecting passively evolving ERGs.

\section{ERG Environments}
Studying the association of ERGs with $K_{s}$-band galaxies and faint radio sources is valuable in understanding their nature.
For this analysis we have used the ERG sample selected with $R - K_{s} > 5$.

Our estimation of the cross-correlation between $K_{s}$-band galaxies and ERGs reveals no significant association.
Given the depth of our survey, in 
which most of the $K_{s}$-band galaxies lie at $z \ll 1$, this is 
not unexpected, since ERGs are found predominantly at $z\approx 1$. 
Georgakakis et al.\ \citeyearpar{Geo2005} find that ERGs are only 
associated with overdensities of $K_{s}$-band galaxies at $z \gtrsim
1$.

\begin{figure}[h]
\centering
\includegraphics[scale=0.5]{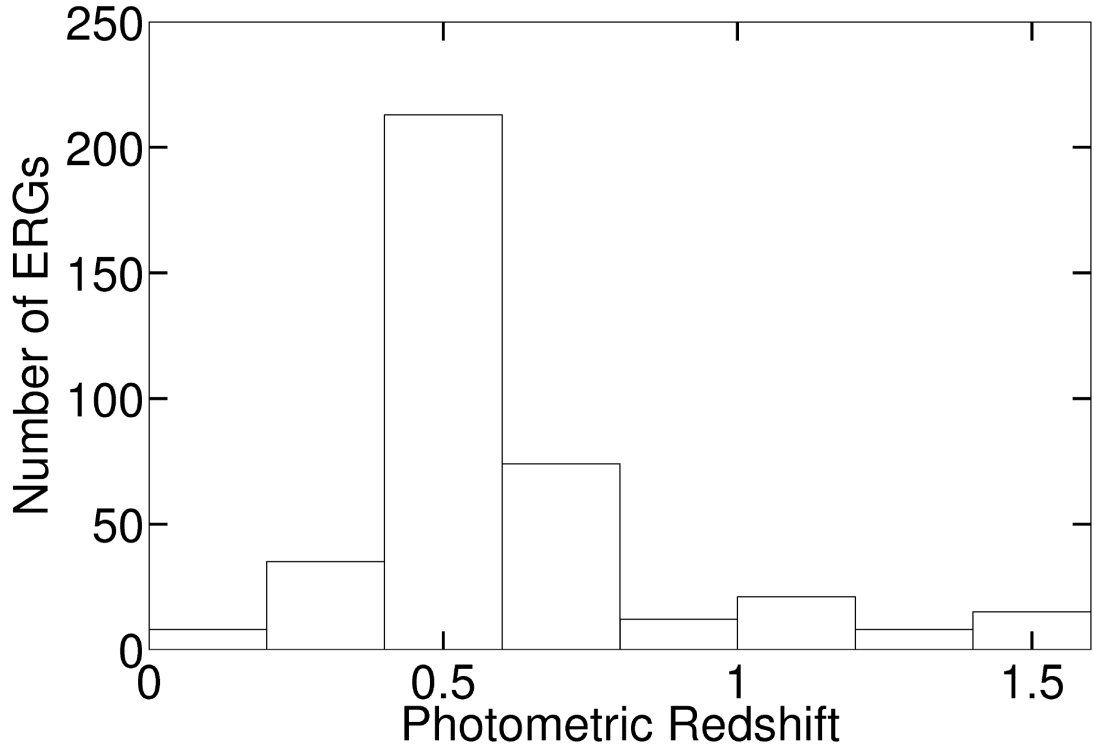}
\caption{Photometric redshift distribution for the $R - K_s>5$ ERG sample.}
\label{ERGpzbincoarse}
\end{figure}

To compare the ERG and faint radio source populations, we first make use of the publicly available code \begin{sc}hyperz\end{sc} (\citealt{Bol2000}) to find photometric redshifts for all galaxies detected in at least three of the {\it UBVRIK$_s$} bands.
In Figure \ref{ERGpzbincoarse} we have shown the photometric redshift distribution for $R - K_s$ ERGs only.
We then match our radio catalogue with the galaxies (including non-ERGs) for which we have photometric redshifts, dividing the radio sources into two broad redshift ranges.
Faint radio sources individually associated with ERGs are excluded, and cross-correlation functions are estimated (e.g. Georgakakis et al.\ \citeyear{Geo2005}) to compare the two radio samples with the ERG catalogue.
Figure \ref{radiozERG}, in which bootstrap errors are shown (Barrow, Bhavsar \& Sonoda\ \citeyear{Bar1984}), demonstrates that while there is no significant 
correlation between the positions of ERGs and faint radio sources
with counterparts at $z_{\rm phot} < 0.4$, there is such a correlation at the 
$2.3\,\sigma$ confidence level for the radio sources with counterparts at $z_{\rm phot} > 0.5$. 
Georgakakis et al.\ \citeyearpar{Geo2005} also detect such a correlation 
at the $\approx 2\,\sigma$ confidence level. 

Our work supports the conclusion that at higher redshifts ($z \gtrsim
0.5$), ERGs and radio sources are associated. This is perhaps not 
surprising since the redshift distributions of both samples peak at 
$z \approx 1$ (Condon \citeyear{Con1989}, McCarthy 
\citeyear{McC2004}). Since we have eliminated faint radio sources individually associated with ERGs,
this cannot simply be the 
result of one-to-one associations. If ERGs and radio sources trace 
galaxy overdensities at higher redshift, we should be able to detect 
strong association between the two populations in high redshift 
cluster environments. In the next section we search for this effect.

\section{Cluster Candidates}
 Rich clusters at high redshifts contain massive, red elliptical 
galaxies at their cores, which can show up as ERG overdensities a 
few arcminutes across in ERG surveys (e.g.\ Stanford et al.\ 
\citeyear{Sta1997}, Roche et al.\ \citeyear{Roc2002}, 
D{\'{\i}}az-S{\'a}nchez et al.\ \citeyear{Dia2007}). Studies of such 
cluster samples may be used to investigate how the highest density 
regions of the universe have evolved, and possibly place constraints 
on their formation.

Here we identify cluster candidates in two different ways. First, we 
use a simple counts-in-cells criterion, and second, we use an 
overdensity-mapping algorithm. The former technique gives us a set 
of likely candidates, and the latter technique allows us to 
treat cluster environments statistically.

\subsection{Counts-In-Cells}
\begin{figure}[h]
\centering
\includegraphics[scale=0.5]{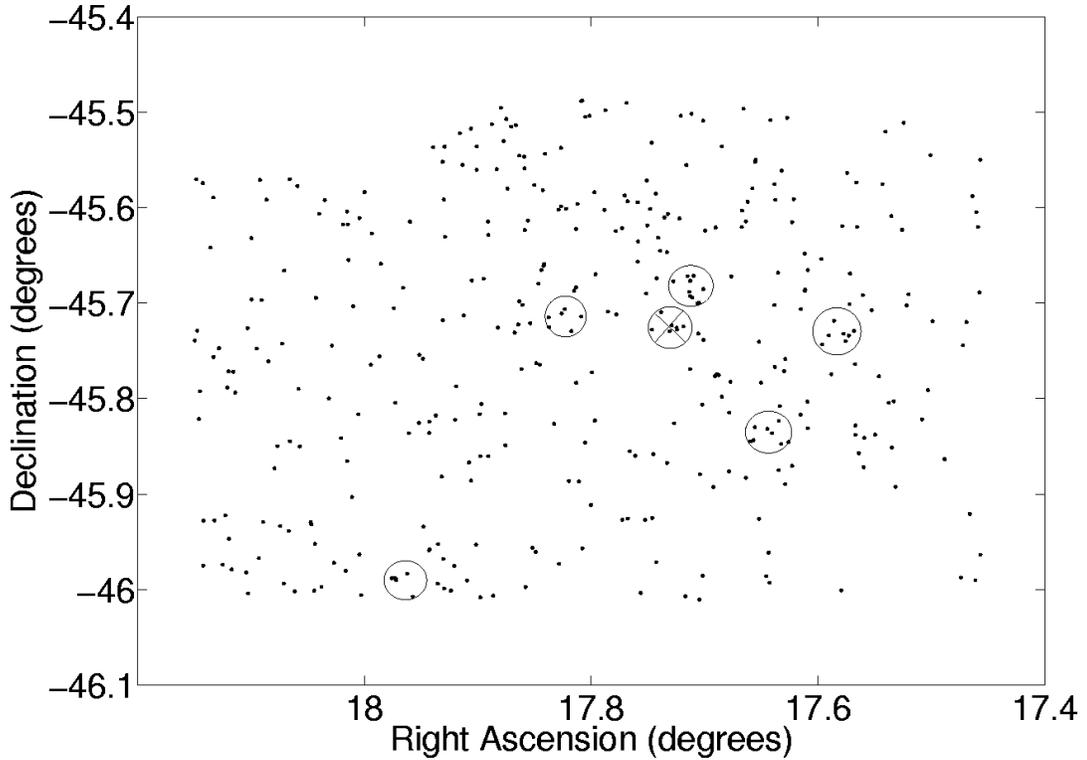}
 \caption{Our ERG field showing the circled locations of the cluster 
candidates based on our counts-in-cells criterion (see section 5.1). 
The cross indicates the probable $0.5 < z < 1$ cluster discussed in 
section 6.}
 \label{sighting}
\end{figure}

\begin{figure*}
\centerline{{\includegraphics[width=7.0cm]{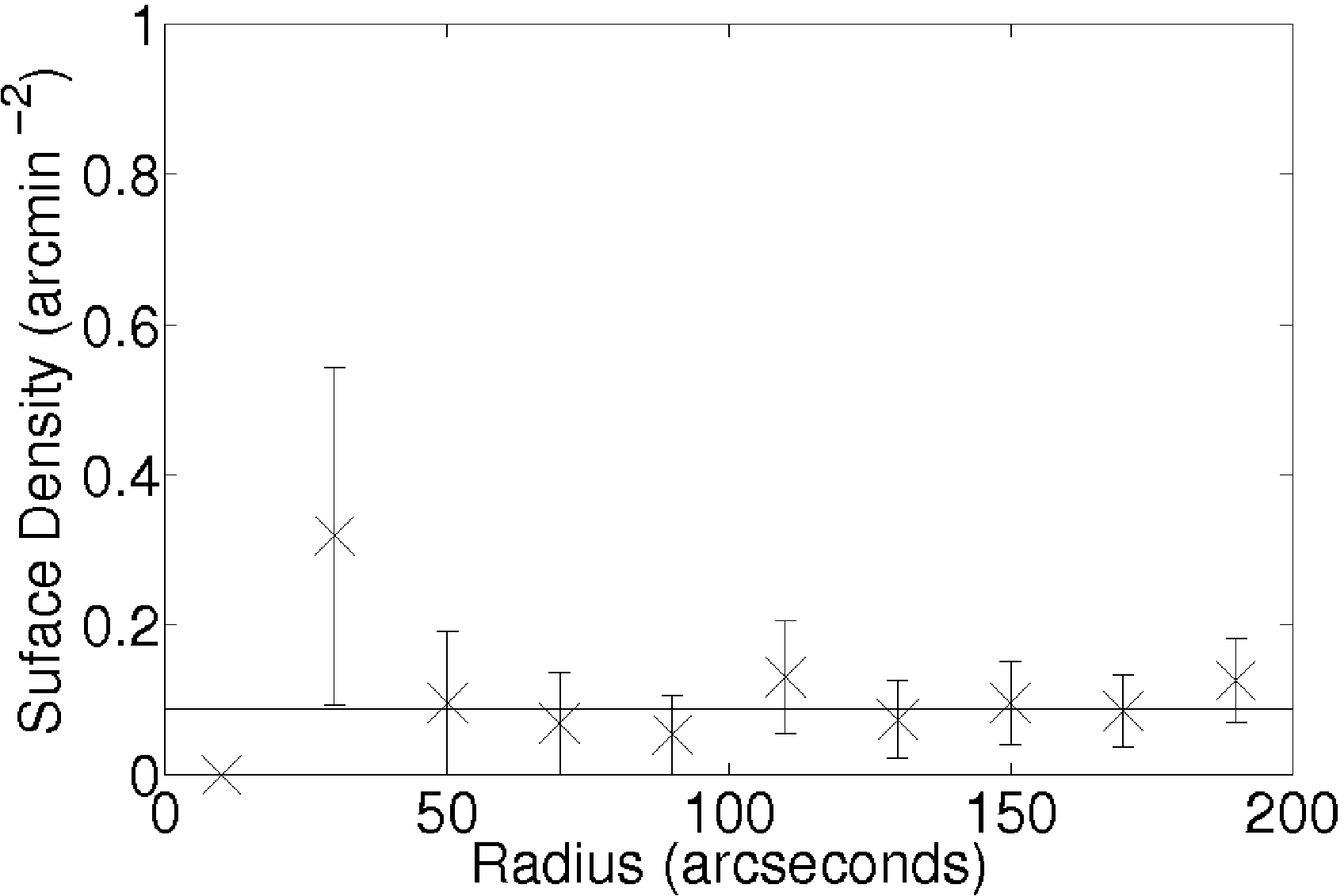}}
{\includegraphics[width=7.0cm]{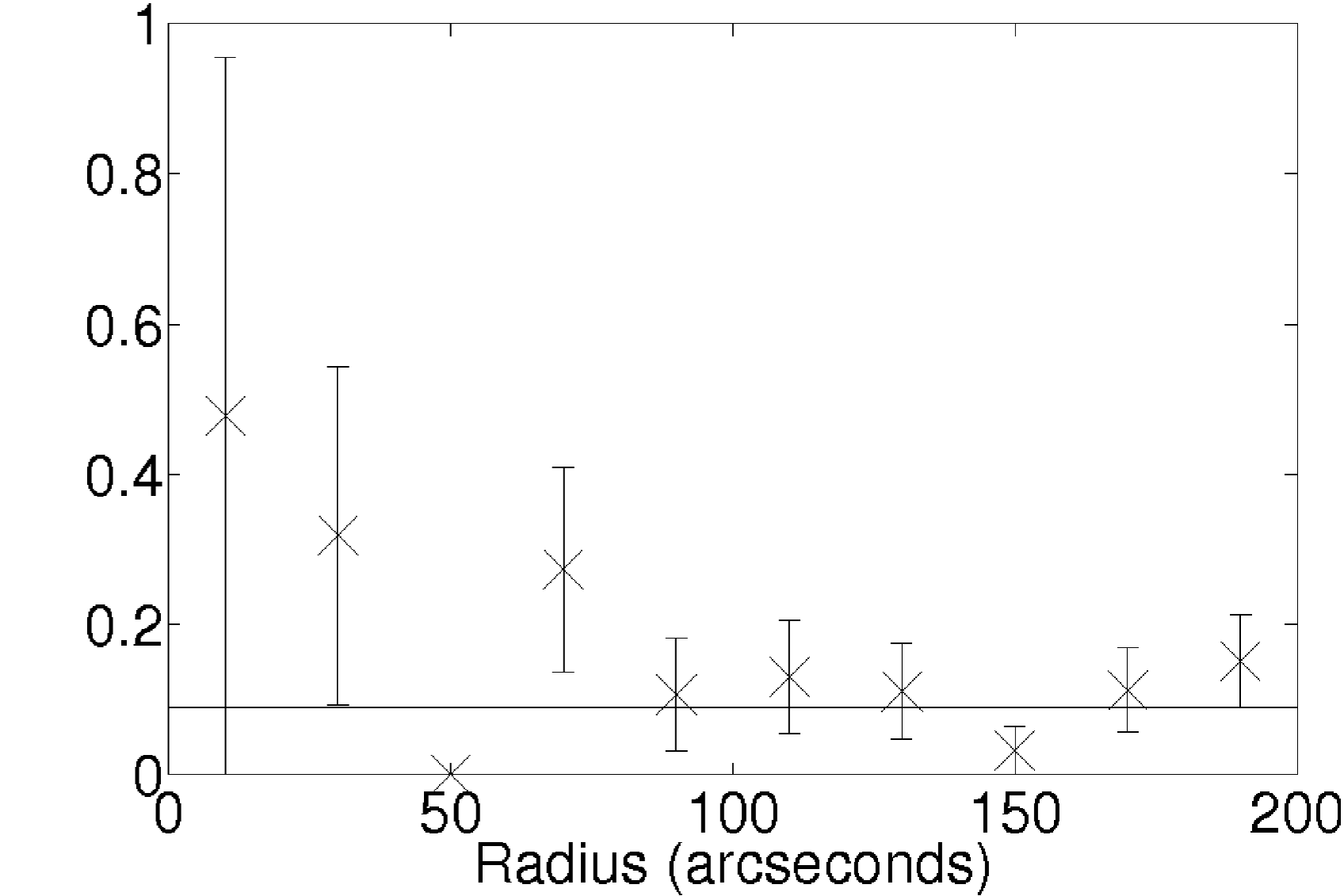}}}
 \caption{Density profiles for faint radio sources around the six 
positions (see Figure \ref{sighting}) which lie at the centres of overdensities selected via the 
counts-in-cells criterion. \textit{Left}: $z_{\rm phot}<0.4$ radio 
sources, and {\it Right:}  $z_{\rm phot}>0.5$  sources. For 
this 
measurement we have not eliminated objects that are both ERGs and 
faint radio sources. The horizontal lines in each panel correspond 
to the average radio source density.}
 \label{radioclusterenvirons}
\end{figure*}

For the entire area of our survey we have found local counts of ERGs within circles of radius 50 arcseconds, and identified six locations (Figure \ref{sighting}) which contain six or more ERGs within a circle of radius 50 arcseconds (see also Blake \& Wall\ \citeyear{Bla2002}).
Given a uniform random field with the same mean density as our ERG catalogue, the probability of a particular location hosting such an overdensity is 3.8 $\times$ 10$^{-4}$.
This criterion selects locations hosting the highest densities and is intended to identify overdensities with the highest probability of physical significance.
D{\'{\i}}az-S{\'a}nchez et al.\ \citeyearpar{Dia2007} report 
one such overdensity as a cluster at $z \approx 1$. Figure 
\ref{radioclusterenvirons} shows the distribution of faint radio 
sources in these cluster environments, for the same redshift ranges 
used in Figure \ref{radiozERG}. One might expect from the 
statistically significant result for $z>0.5$ in Figure 
\ref{radiozERG} that there should also be an excess of associated 
radio sources in the regions with ERG overdensities.  However, we 
find only a 1.05\,$\sigma$ excess of faint radio sources in these 
putative clusters.  This may be because the ERGs in these high 
density regions are pERGs which have fewer radio counterparts 
since they have lower star-formation rates.


\subsection{Overdensity-Mapping Algorithm}
We now describe an algorithm we have developed to identify overdensities of ERGs within our survey area, beginning with a general description before explaining the procedure in detail.
For any given location, we compare the local galaxy count with the distribution of counts expected from a large set of uniform random catalogues, each with the same mean density as the ERG catalogue.
This comparison allows us to calculate the probability that the local area contains a greater number of galaxies than that expected from a random distribution.
Repeating this process for a range of specific radii, we form a \textit{probability function} which is then associated with the location.
Similarly, a probability function is assigned to each of a set of uniformly spaced locations within our survey.
The optimal range and resolution of the radii and locations sampled may be found empirically.

Each probability function is used to estimate the probability of the count within each radius enclosing its location occurring as the result of a uniform random distribution.
If the probability of obtaining the local count from a random catalogue is high, the value of the probability function will be close to a half, since the number of random catalogues with lower counts will be similar to the number with higher counts.
In the presence of a genuine overdensity, the probability function will rise above 0.5 and continue to ascend with increasing radius (since the probability of a large overdensity is lower than the probability of a small one) until the ``edge" of the overdensity has been reached.
Beyond this edge, the enclosed density will fall back to the background density, and the probability function will approach 0.5.
Hence, we expect each probability function to be sensitive to both the significance and scale of any nearby overdensities, characterised by its maximum and the radius at which it occurs.

For overdensities centred on particular catalogue objects, this maximum occurs at a radius of zero (which thus encloses an infinite density), but we negate this by specifying a minimum overdensity membership of two.
We define the radius at which the maximum occurs as the \textit{overdensity radius}, and the \textit{overdensity probability} as the value which the probability function takes at the overdensity radius.
Our technique uses probability to compare between scales in order to select the most likely scale at which each particular location hosts an overdensity.

To achieve this, we follow these steps.
\begin{itemize}
\item The survey area is covered with a uniform grid of test points at intervals of 20 arcseconds along both right ascension and declination.
\item We populate the survey area with a large number of random sets, each containing the same average density of points as the ERG catalogue that we are examining.
\item We count the number of ERGs and random objects (in each random set) as a function of distance from each test point, in steps of 10 arcseconds from 20 to 120 arcseconds.
\item The probability of overdensity is then the fraction of random sets for which the number of ERG counts exceeds the number of random counts. ERG counts consistent with uniform random distribution will therefore produce overdensity probabilities of 0.5.
\end{itemize}
In overdense regions, the overdensity probability directly quantifies the probability that the local environment is associated with an overdensity, and the overdensity radius directly characterizes the scale size of this structure.
Each of the grid points which we test within our catalogue will have an associated overdensity probability and overdensity radius.
By confining our random sets to the area of the catalogue under inspection, we account for edge effects.

The probability of overdensity is intimately connected with the overdensity radius.
We are effectively sensitive to multi-scale structure throughout the survey area, meaning that we will detect both high density small-scale features (which might be overlooked in the case of simple large-scale smoothing), and extended regions of intermediate density (which might be overlooked in the case of simple small-scale smoothing).
Our algorithm is therefore sensitive to overdensities spanning a wide range of scales.
In particular it has the potential to simultaneously identify filamentary structures as well as clusters, an aspect that we will explore in detail in future work.

\begin{figure}[h]
\centering
\includegraphics[scale=0.6]{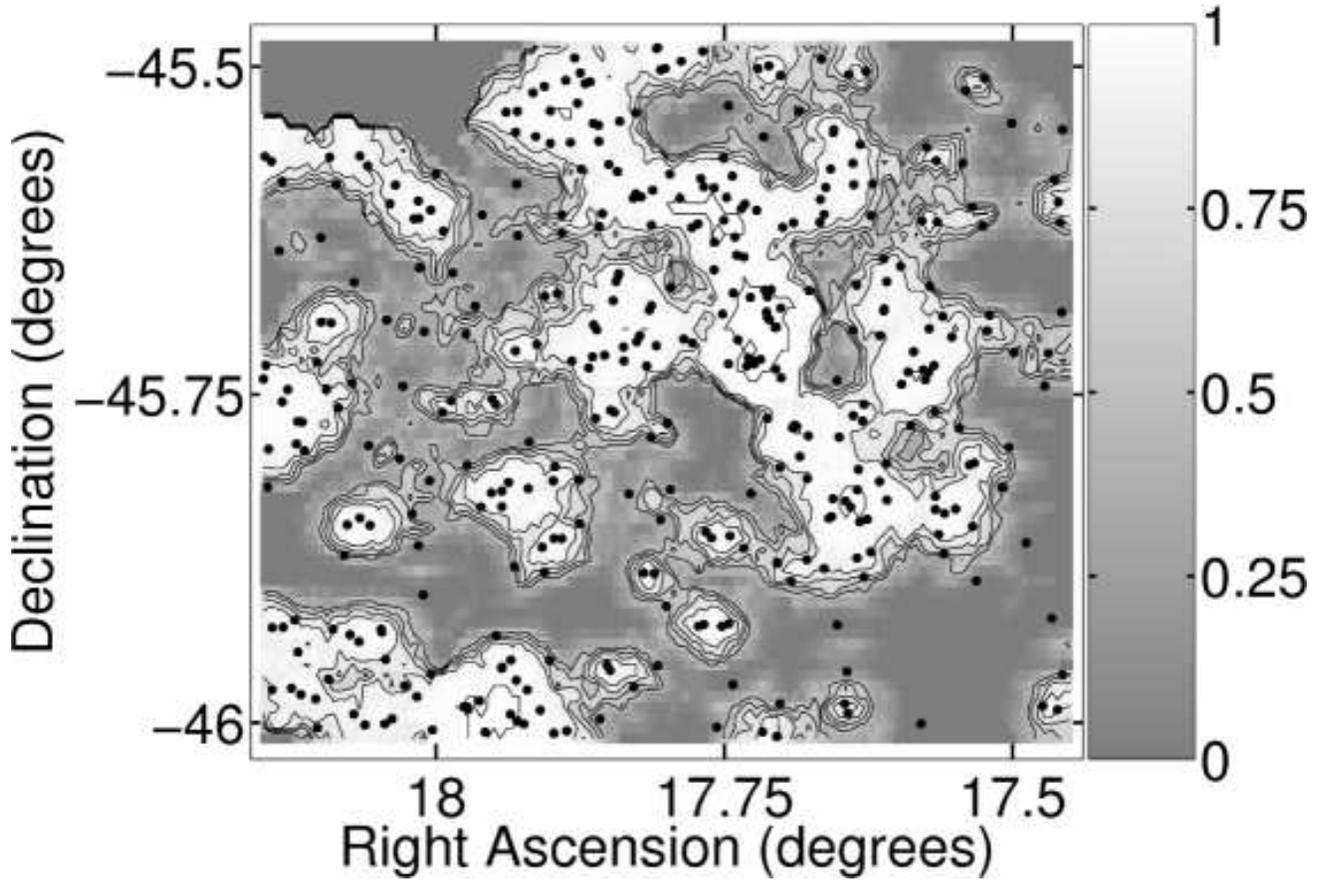}
 \caption{Grayscale representation of our overdensity probability 
map for the $R - K_{s}$-selected ERG sample (see section 5.2). The 
black dots are the ERGs in this sample and the contours correspond 
to probabilities of 0.5, 0.625, 0.75, 0.875 and 1.  There is clear 
evidence of filamentary structure in the high density islands.}
 \label{RKssmoothRKsERGs}
\end{figure}

Figure \ref{RKssmoothRKsERGs} shows the result of the overdensity-mapping algorithm applied to our data.
It is clear that the regions identified by our algorithm as having a high overdensity probability do indeed correspond to visual overdensities of ERGs, and similarly that regions with no ERGs, or isolated ERGs, have low probability.

\subsection{Faint Radio Sources and the Overdensity Map}
\begin{figure}
\centering
\includegraphics[scale=0.5]{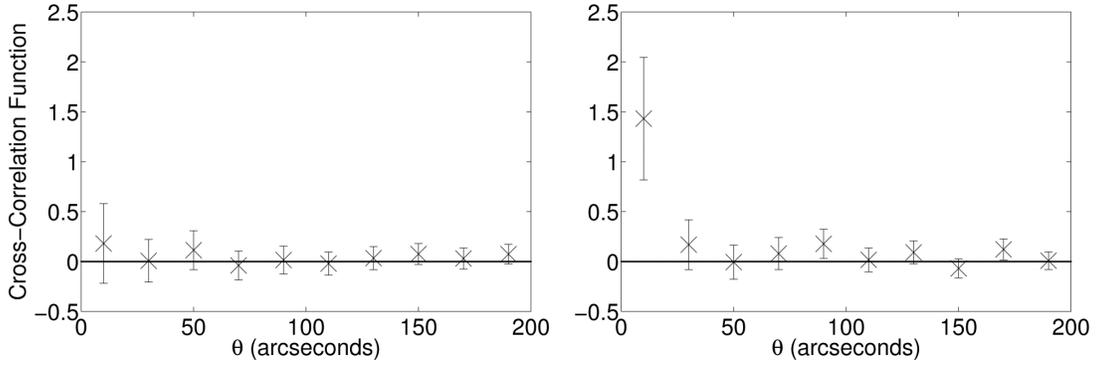}
\caption{Cross-correlation functions with bootstrap errors for ERGs around faint radio sources with optical counterparts assigned
photometric redshifts (not shown in Figure \ref{ERGpzbincoarse}). {\it Left:} 97 radio sources associated with galaxies at $z_{\rm 
phot} < 0.4$ and {\it Right:} 98 sources with $z_{\rm phot} > 0.5$.
As discussed, our exclusion of faint radio sources individually associated with ERGs ensures that the small-scale feature is not the result of one-to-one associations.}
 \label{radiozERG}
\end{figure}

\begin{figure*}
\centerline{{\includegraphics[width=7.0cm]{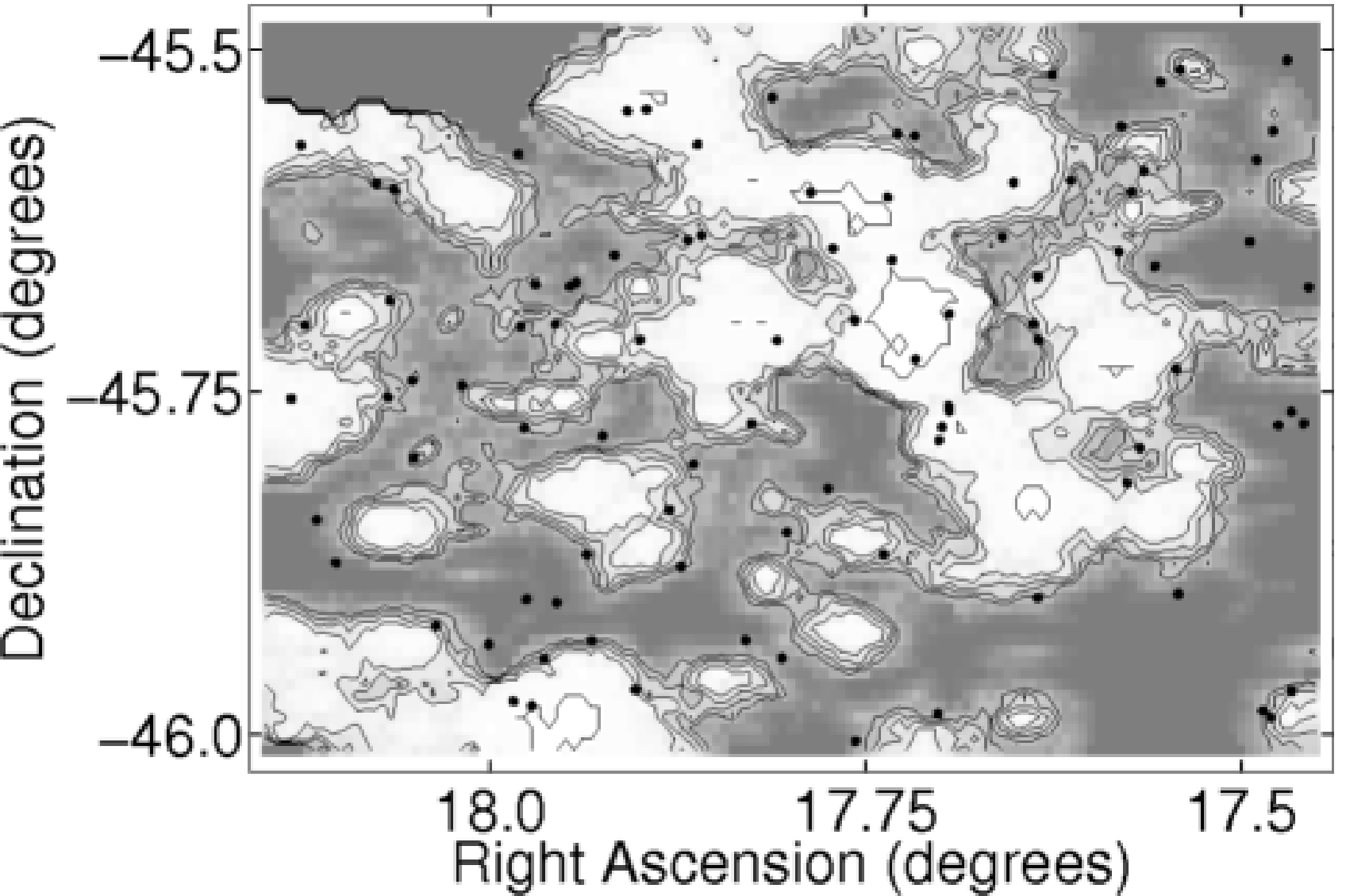}}
{\includegraphics[width=7.0cm]{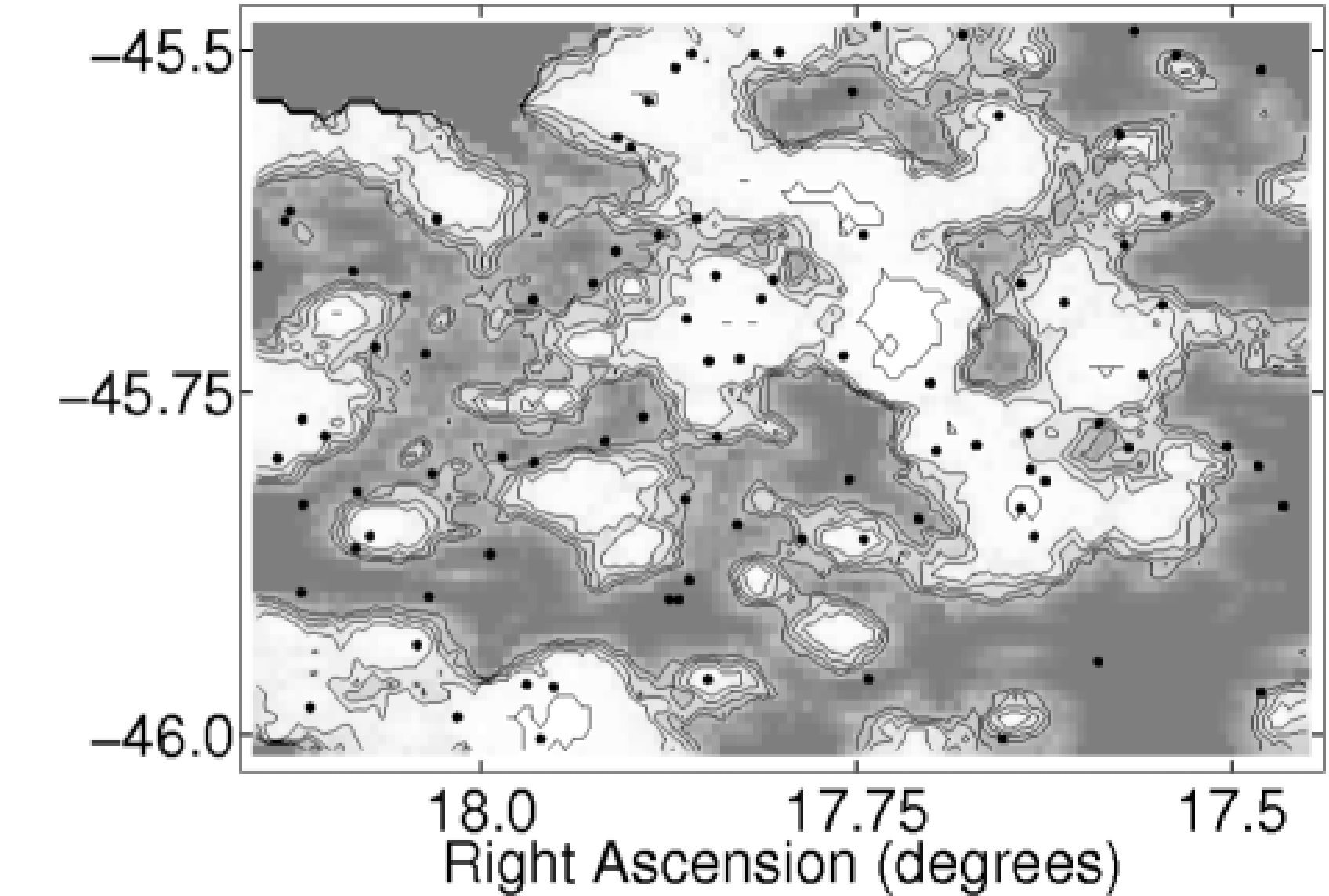}}}
 \caption{Grayscale as in Figure \ref{RKssmoothRKsERGs}, with black dots indicating the 
locations of faint radio sources. \textit{Left:} 97 sources with 
$z_{\rm phot} < 0.4$. \textit{Right:} 98 sources with $z_{\rm phot} 
> 0.5$.}
 \label{RKssmoothzradio}
\end{figure*}

In Figure \ref{RKssmoothzradio} we visually compare the overdensity probability map in Figure \ref{RKssmoothRKsERGs} with the distribution of faint radio sources in the two $z_{\rm phot}$ ranges previously defined.
The redshift dependence shown in Figure \ref{radiozERG} is not apparent in Figure \ref{RKssmoothzradio}.
We stress, however, that since our algorithm has smoothed the predominantly $z \approx 1$ ERG population, the overdensity map will be biased towards structures at this redshift.

\begin{figure*}
\centerline{{\includegraphics[width=7.0cm]{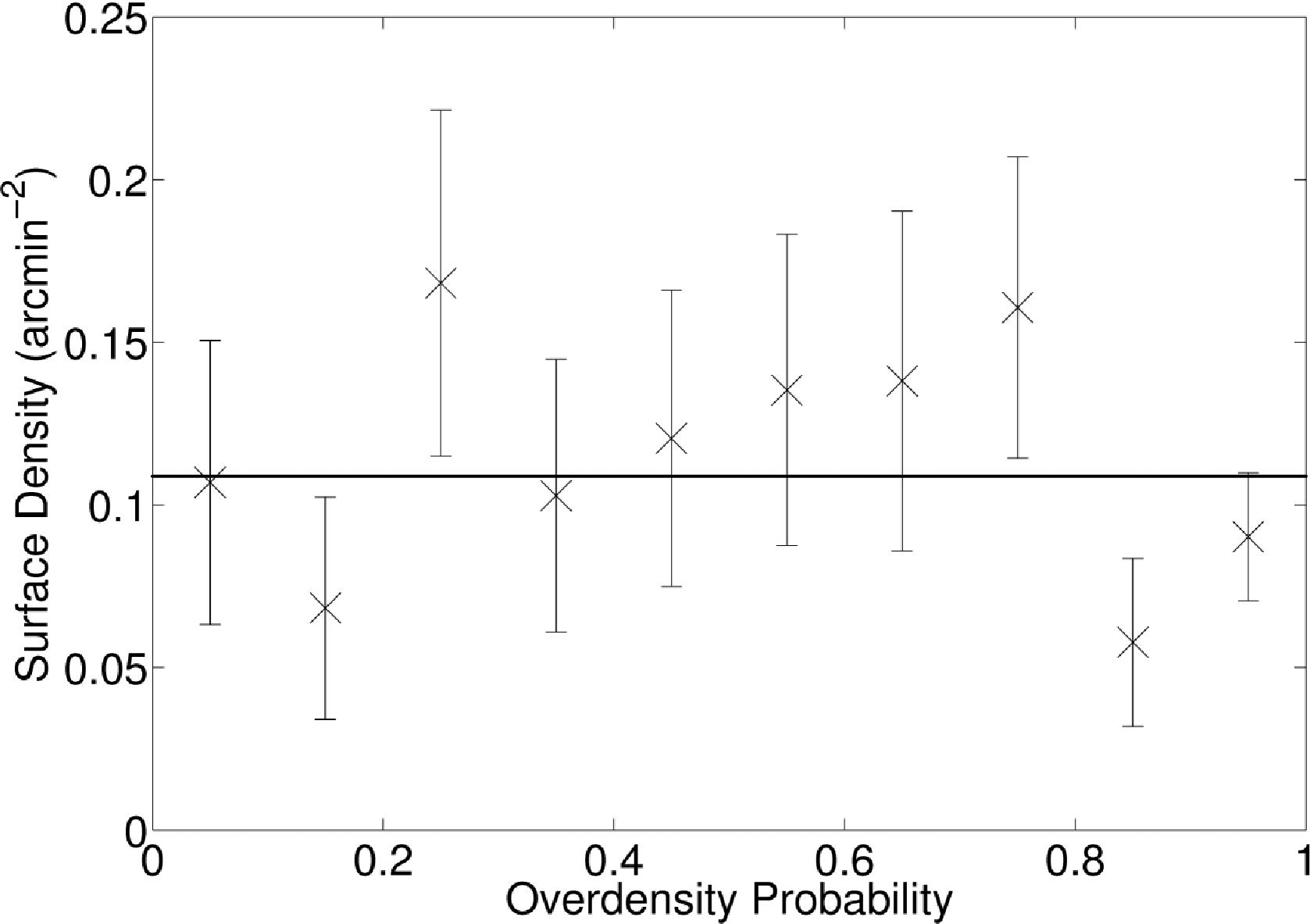}}
{\includegraphics[width=7.0cm]{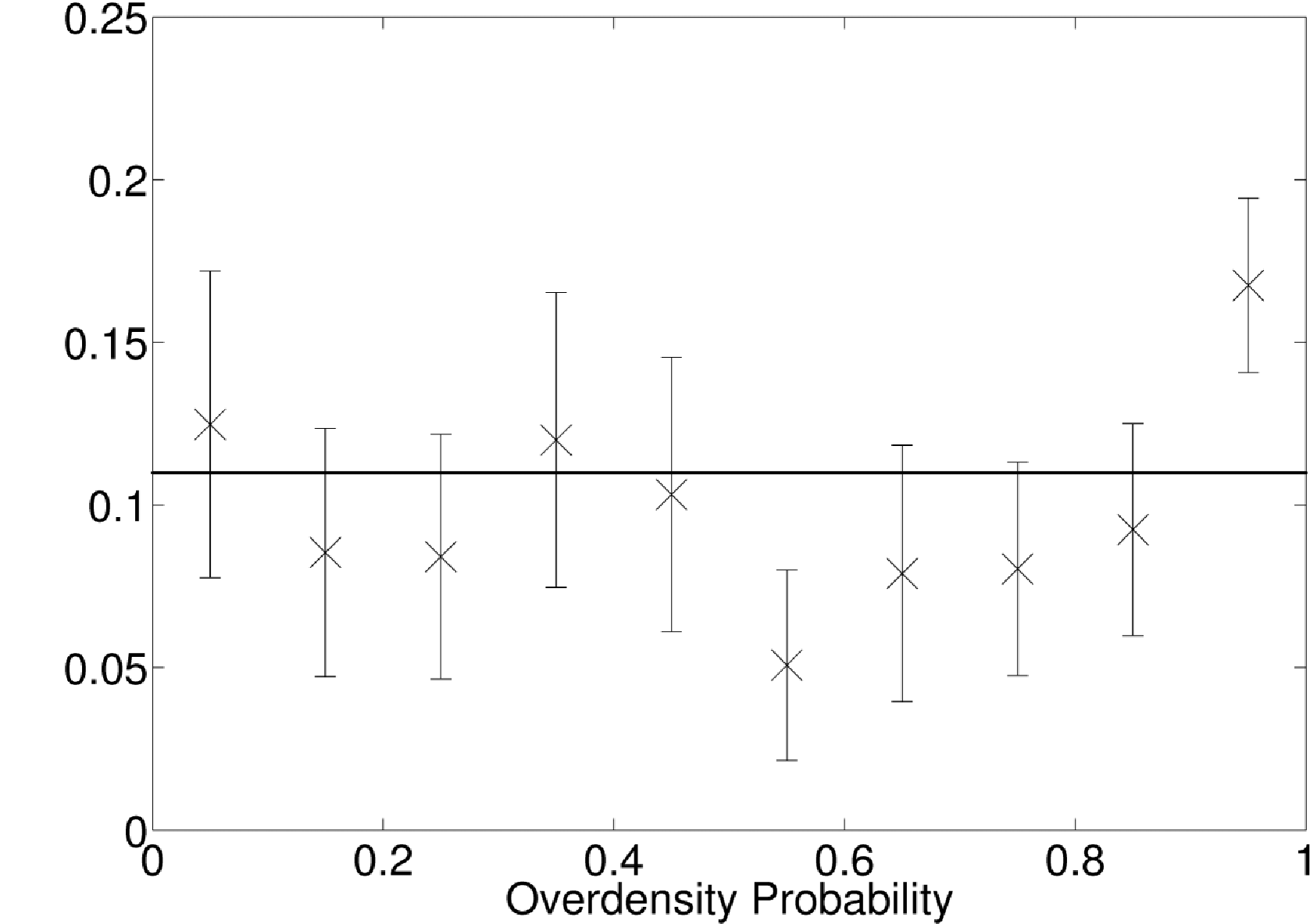}}}
 \caption{Surface density of radio sources as a function of 
overdensity probability.  \textit{Left:} Sources with $z_{\rm phot} 
< 0.4$. \textit{Right:} Sources with $z_{\rm phot} > 0.5$. The error 
bars are Poissonian and the horizontal lines correspond to the 
average source density.}
 \label{smoothhistpzradio}
\end{figure*}

We have matched each of the faint radio sources with the pixel (from our overdensity map) closest to its position, and counted the frequency with which faint radio sources are matched with certain pixel values.
Effectively, what we are doing here is counting the number of faint radio sources which lie along different probability contours of our overdensity map.
The result of this analysis is shown in Figure \ref{smoothhistpzradio}, which suggests that faint radio sources with optical counterparts with $z_{\rm phot} > 0.5$ are indeed found in high density regions, with a 2$\sigma$ excess for overdensity probabilities between 0.9 and 1.
This is not seen in the $z_{\rm phot} < 0.4$ sample.
While the statistics are poor, and our overdensity map is strongly biased towards higher redshifts (as shown in Figure \ref{ERGpzbincoarse}), Figure \ref{smoothhistpzradio} tentatively suggests that clusters at high redshift are more likely to host faint radio sources in their central regions.

The lack of a clear association of ERGs and radio sources with optical counterparts with $z_{\rm phot} > 0.5$ evident in Figure \ref{RKssmoothzradio} and the result from Figure \ref{radioclusterenvirons} suggests that the correlation between ERGs and moderately high redshift radio sources implied by Figure \ref{radiozERG} does not occur in regions with the highest ERG densities.
Figure \ref{radiozERG} shows association on scales which may not be noticeable when considering the scale of the entire field shown in Figure \ref{RKssmoothzradio}, and cannot be used to draw conclusions about the characteristic density of the environment in which ERGs and faint radio sources are associated.
Figure \ref{smoothhistpzradio} illustrates, however, that our overdensity map provides a way of directly probing this characteristic density.

\section{A Possible Cluster at $0.5 < z < 1$}
 Of the cluster candidates identified via the counts-in-cells 
criterion (Figure \ref{sighting}), we have singled out one in 
particular as an interesting object for future study.

\begin{figure*}
\centerline{{\includegraphics[width=7.0cm]{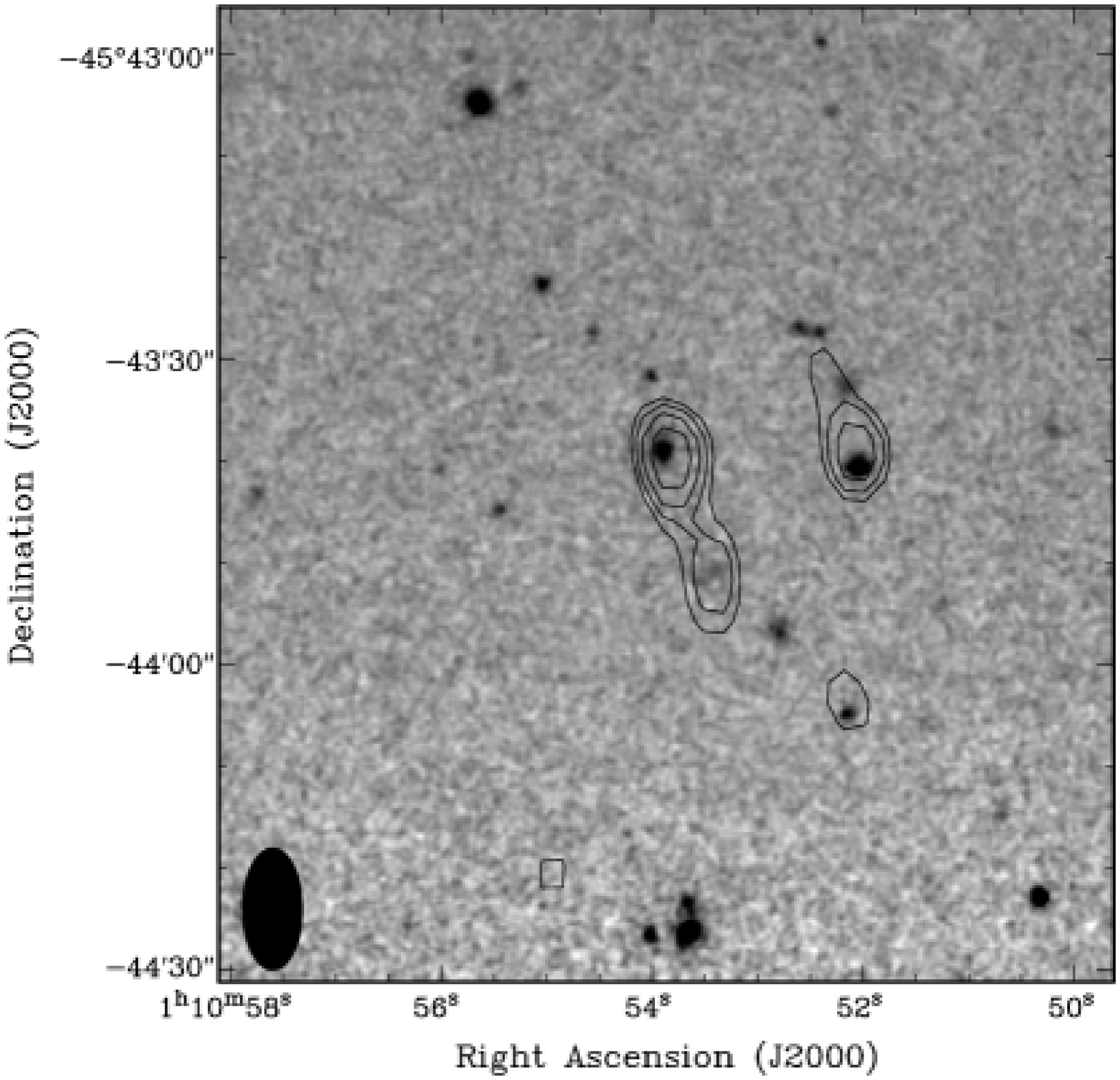}}
{\includegraphics[width=7.0cm]{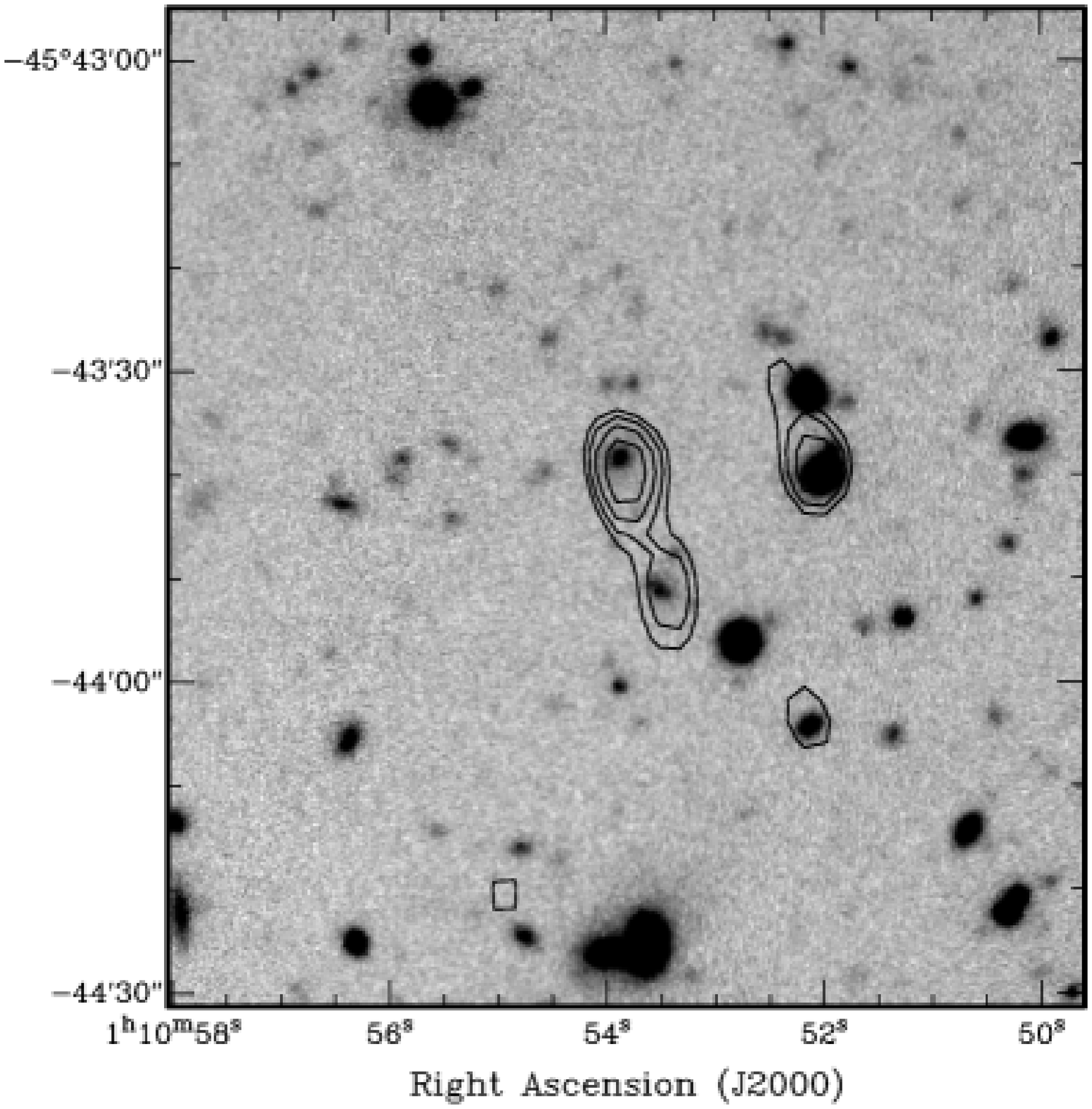}}}
 \caption{Images of a possible $0.5 < z < 1$ cluster 
(see Section 6).  {\it Left:} $K_{s}$-band; {\it Right:} 
$R$-band.  Overlaid contours at 1.4 GHz from the Phoenix 
Deep Survey are at 36, 47, 61 and 79 $\mu$Jy/beam.
The ATCA synthesised beam (12$^{\prime\prime}\times 6^{\prime\prime}$ in position angle 0$^{\circ}$) is shown in the lower left hand corner of the $K_{s}$-band image.
}
 \label{clusterpics}
\end{figure*}

$K_{s}$-band and $R$-band images of 
this putative cluster are shown in Figure \ref{clusterpics}, with ATCA 1.4\,GHz contours overlaid.
Astrometry for both images was referenced to 8--10 stars in the 
SuperCOSMOS (Hambly et al.\ \citeyear{Ham2001}) $R$-band image which has formal fit errors of 0.11 
arcsec rms in each coordinate. The bright ($K_{s} = 16.3$) central 
radio source PDF\,J011053.8$-$454339 has a flux density of $S_{1.4} =$
$106\,\mu$Jy. At this flux density, the relative proportions of 
starburst galaxies and AGNs in the faint radio population are 
strongly model dependent (e.g. Seymour et al. \citeyear{Sey2004}, Hopkins \citeyear{Hop2004}), although recent observations (Seymour et al. \citeyear{Sey2008})
indicate that star-forming galaxies at this flux
density are more numerous than AGNs by a factor of 2 to 3. Figure $\ref{clusterpics}$ shows that the central radio source
has a nearby companion radio source, PDF\,J011053.4$-$454351, 
associated with a faint $K_s$-band object that falls below our 
catalogue threshold.

\begin{figure}[h]
\centering
\includegraphics[scale=0.5]{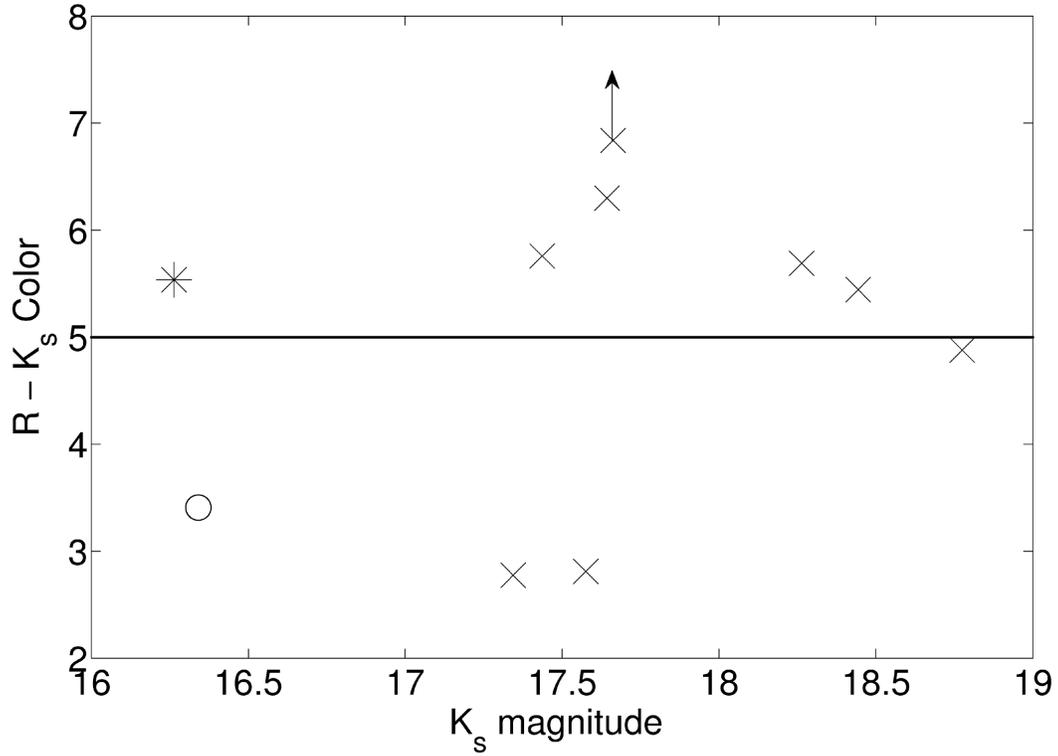}
 \caption{Colour-magnitude diagram for the galaxies in the probable 
$0.5 < z < 1$ cluster (Figures \ref{sighting} and \ref{clusterpics}) within $22''$ of the brightest 
galaxy. The brightest galaxy is marked with an asterisk, and the western tailed radio source host is marked with a circle.}
 \label{clustercolourmag}
\end{figure}

Figure \ref{clustercolourmag} shows there are six objects in this 
possible cluster with $R - K_{s} \ge 5.4$.  Since such extreme 
colours are caused primarily by redshift, the cluster is likely to 
lie at $0.5 < z < 1$ but not much greater, given the depth 
of our $K_s$ survey. If it lies at $z=1$, the radio luminosity of the 
central ERG is $L_{1.4}=5\times 10^{23}\,$W\,Hz$^{-1}$, making it a 
low-luminosity AGN. As the brightest galaxy in this field, it is 
also consistent with being the central dominant cluster galaxy.

An 80 $\mu$Jy radio source to
the west (PDF\,J011052.0$-$454338) is also shown in Figure $\ref{clusterpics}$, with contours suggestive of a
head-tail morphology. Tailed radio galaxies are often seen in galaxy 
clusters (e.g.\ Gomez et al.\ \citeyear{Gom1997}, Sakelliou \&
Merrifield \citeyear{Sak2000}, Klamer et al.\ \citeyear{Kla2004}),
but its $R=20.4$ magnitude and clear separation from the red locus 
in Figure \ref{clustercolourmag} point to it belonging to the 
foreground cluster evident in the $R$-band image.

\begin{figure}[h]
\centering
\includegraphics[scale=0.5]{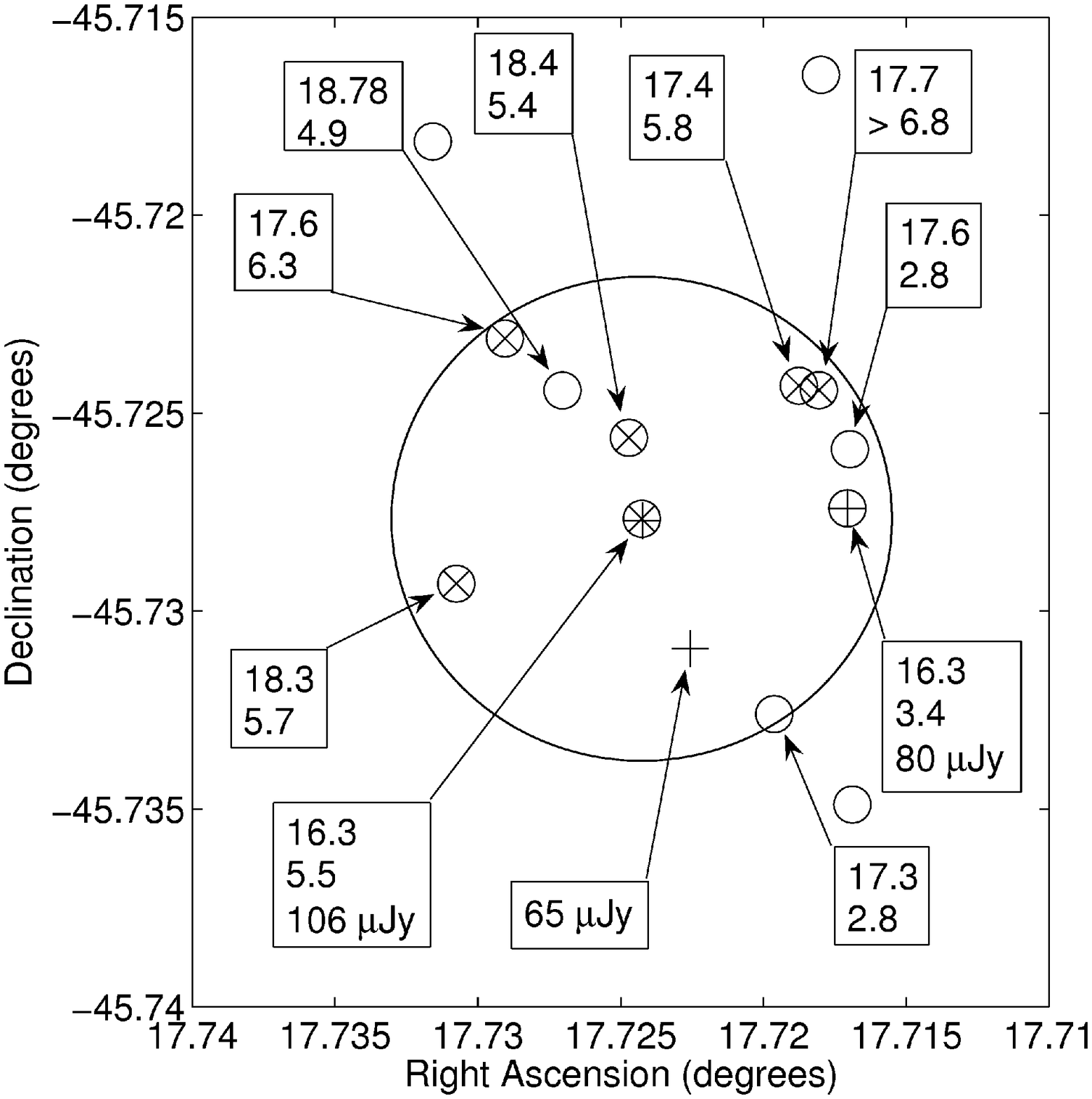}
 \caption{The vicinity of the probable $0.5 < z < 1$ cluster, showing 
the locations of cataloged $K_{s}$-band galaxies ({$\bigcirc$}), 
faint radio sources ({$+$}) and ERGs ({$\times$}). Where available, 
we have labeled each object with $K_{s}$ magnitude, $R - K_{s}$ 
colour and radio flux density.
A circle of radius 22 arcseconds is also shown.}
 \label{clearviewwithcircleandarrows}
\end{figure}

Figure \ref{clearviewwithcircleandarrows} shows that the ERGs 
associated with this overdensity all fall within 22 arcseconds of 
the central ERG, corresponding to a projected radius of $\sim$0.18 
Mpc at $z \approx 1$.  This is less than 1/5 of an Abell radius 
(e.g.\ \citealt{Abe1965}). However, the radial extent of the cluster 
is likely to be underestimated because of our relatively bright 
$K_s$-band limiting magnitude.
The presence of radio sources within our empirically determined radial extent suggests that this estimate is consistent with the core radius of 167 kpc found for nearby clusters, based on their radio source distribution (Ledlow \& Owen\ \citeyear{Led1995}).


\section{Summary and Future Work}
Within the
Phoenix Deep Survey we have identified a sample of ERGs over a large area (1160 arcmin$^2$)
that is 80\% complete to a limiting magnitude of 
$K_{s} = 18.5$. The number counts and clustering properties of our 
ERG sample are consistent with previous observations. Based on photometric redshifts we find 
evidence for the association of ERGs and faint radio sources with
$z_{\rm phot} >0.5$, but not with $z_{\rm phot} < 0.4$, consistent 
with earlier results on the evolution of star formation and AGN 
activity.


We find weak evidence to suggest that $R - K_{s}$-selected ERGs are 
more strongly clustered than $I - K_{s}$-selected ERGs. If confirmed 
with more extensive datasets, this would imply that $R - K_{s}$-selected ERGs are more strongly associated with overdensities, 
supporting the suggestion by V\"ais\"anen \& Johansson 
\citeyearpar{Vai2004} that $R - K_{s}$-selected ERGs contain a 
higher proportion of passively evolving sources, assuming that the 
two ERG populations have similar redshift distributions. 

The identification of the cluster candidates in this study shows 
that ERGs can be used to identify overdensities of potential physical significance, and our 2.3\,$\sigma$ 
result for the association of ERGs and faint radio sources at 
$z_{\rm phot} > 0.5$ is evidence that both classes of objects trace 
overdense regions.
As passive galaxies dominate in high density environments, the absence of radio source overdensities in most of our cluster candidates supports the results of Georgakakis et al. \citeyearpar{Geo2006} and Simpson et al. \citeyearpar{Sim2006} that dsfERGs are more closely associated with the radio population than pERGs.
As a result, faint radio sources are most efficient at identifying 
starburst galaxies in the infall regions of clusters. The probable 
$0.5 < z < 1$ cluster identified here will be explored in more 
detail with spectra and deeper imaging.


We have introduced a method for quantifying overdensities in a 
galaxy distribution that is sensitive to overdensities on a broad 
range of scales, and demonstrated its efficiency in identifying 
cluster candidates.
A faster implementation of this algorithm is being developed.


ERG sample size is a strong function of limiting $K_s$-band 
magnitude, and it is necessary to have complementary deep optical 
$R$- and $I$-band data to detect ERGs with the most extreme colours. 
Large samples of ERGs are becoming available with the next generation 
of large area near-infrared surveys, such as UKIDSS
(Lawrence et al. \citeyear{Law2007}). Where these overlap with existing deep 
radio surveys, it will be possible to explore the association of the 
faint radio population with ERGs in unprecedented detail.

\acknowledgements
We thank Helen Johnston, Scott Croom and Chris Miller for their advice with \begin{sc}hyperz\end{sc}, correlation analysis and cluster detection algorithms respectively.
The anonymous referee's comments helped us to improve many aspects of this paper.
AMH acknowledges support from the Australian Research 
Council through a QEII Fellowship (DP0557850).
JA gratefully acknowledges the support from the Science and Technology Foundation (FCT, Portugal) through the research grants POCI/CTE-AST/58027/2004 and PPCDT/CTE-AST/58027/2004.
This research has made use of NASA's Astrophysics Data System.

\clearpage

\begin{deluxetable}{ccccc}
\tablewidth{0pt}
\tablecaption{Correlation Function Amplitudes $A$
\label{clusteringamplitudetable}}
\tablehead{
\colhead{}&&\multicolumn{3}{c}{$K_{s}$ limiting magnitude}\\
\colhead{Selection}&Properties\tablenotemark{a}&\colhead{18.0}&\colhead{18.25}&\colhead{18.5}
}
\startdata
$R - K_{s}$&$A$ ($\times 10^{-3}$)&$28.7 \pm 4.1$&$19.6 \pm 2.8$&$15.0 \pm 2.1$\\
&$N$&143&202&273\\
\hline\\[-3mm]
$I - K_{s}$&$A$ ($\times 10^{-3}$)&$26.9 \pm 4.4$&$16.2 \pm 3.0$&$9.6 \pm 2.2$\\
&$N$&132&188&256
\enddata
\tablenotetext{a}{Uncertainties in $A$ are calculated from 1 $\sigma$ errors in the fitted amplitudes.
The number $N$ of objects brighter than each limiting magnitude is also shown.}
\end{deluxetable}

\clearpage

\end{document}